\documentclass[preprint2]{aastex}
\begin{document}

\title{The Old, Super-Metal-Rich Open Cluster, NGC 6791 - Elemental Abundances
in Turn-off Stars from Keck/HIRES Spectra}

\author{Ann Merchant Boesgaard\altaffilmark{1}, Michael
G. Lum\altaffilmark{1}}

\affil{Institute for
Astronomy, University of Hawai`i at M\-anoa, \\ 2680 Woodlawn Drive, Honolulu,
HI{\ \ } 96822 \\ } 

\author{Constantine P.~Deliyannis\altaffilmark{1}}
\affil{Department of Astronomy, Indiana University\\
727 East 3rd Street, Swain Hall West 319, Bloomington, IN {\ \ }47405-7105}

\email{boes@ifa.hawaii.edu}
\email{mikelum@ifa.hawaii.edu}
\email{cdeliyan@indiana.edu}

\altaffiltext{1}{Visiting Astronomer, W.~M.~Keck Observatory jointly operated
by the California Institute of Technology and the University of California.}

\slugcomment{Accepted by Ap.J.~Dec.~12, 2014}

\begin{abstract}
The study of star clusters has advanced our understanding of stellar
evolution, Galactic chemical evolution and nucleosynthesis.  Here we
investigate the composition of turn-off stars in the intriguing open cluster,
NGC 6791, which is old, but super-metal-rich with high-resolution ($R$ =
46,000) Keck/HIRES spectra.  We find [Fe/H] = +0.30 $\pm$0.02 from
measurements of some 40 unblended, unsaturated lines of both Fe I and Fe II in
eight turn-off stars.  Our O abundances come from the O I triplet near 7774
\AA\ and we perform a differential analysis relative to the Sun from our Lunar
spectrum also obtained with Keck/HIRES.  The O results are corrected for small
nLTE effects.  We find consistent ratios of [O/Fe]$_{\rm n}$ with a mean of
$-$0.06 $\pm$0.02.  This is low with respect to field stars that are also both
old and metal-rich and continue the trend of decreasing [O/Fe] with increasing
[Fe/H].  The small range in our oxygen abundances is consistent with a single
population of stars.  Our results for the alpha elements [Mg/Fe], [Si/Fe],
[Ca/Fe], and [Ti/Fe] are near solar and compare well with those of the old,
metal-rich field stars.  The two Fe-peak elements, Cr and Ni, are consistent
with Fe.  These turn-off-star abundances provide benchmark abundances to
investigate whether there are any observable abundance differences with the
giants that might arise from nuclear-burning and dredge-up processes.
Determinations of upper limits were found for Li by spectrum synthesis and are
consistent with the upper limits in similar stars in the relatively old,
super-metal-rich cluster NGC 6253.  Our results support the prediction from
standard theory that higher-metallicity stars deplete more Li.  Probably no
stars in NGC 6791 have retained their initial Li.
\end{abstract}


\section{INTRODUCTION}

The study of both open clusters and globular clusters have contributed to our
understanding of the formation and chemical evolution of the Galaxy and the
evolution of stars.  The stars in open clusters are thought to have formed at
the same time, within a few megayears, thus comparisons of clusters of
different ages have revealed some details of the evolution of stars.  Ages
have been determined by the turn-off point from the main sequence with
theoretical predictions from isochrones.

A given open cluster is formed from a giant molecular cloud thus they begin
with the same composition and same space motion.  Open clusters span the age
of the Galaxy, but old open clusters are rare.  Typical open clusters have
some tens to hundreds of stars and as they move through the Galaxy, they tend
to dissipate.  Their former members become field stars.  The oldest clusters
survive only if they are massive and dense.  Examples are NGC 188, NGC 6253,
and NGC 6791.

Almost every field in astrophysics has benefited in some way from studies of
star clusters.  The study of both open clusters and globular clusters have
especially contributed to our understanding of the formation and evolution of
the Milky Way Galaxy, and of the evolution of stars.  Since we can age-date
clusters, we can learn about age-dependent phenomena.  The distributions of
stars in color--magnitude diagrams (CMDs) and the inferred cluster ages have
confirmed many of the strictures of the standard stellar evolution theory,
which, in turn, is used in the construction of galactic population synthesis
and chemical evolution models.  For example, stellar evolution theory explains
why most stars are found on the main sequence and why they spend so much time
there.  Even long ago, it had been inferred from CMDs that stars evolve from
the main sequence to become red giants, though how they did this was a mystery
for a long time.  That is, until stellar evolution theory provided the much
anticipated understanding.  One of the biggest triumphs of stellar evolution
theory from relatively recent years is the prediction that most globular
clusters are about 12--13 Gyr old, just younger than the Planck-inferred age
of 13.8 Gyr of the universe (Ade et al.~2014).  In spite of these and other
numerous successes, standard theory, nevertheless, fails to explain the vast
majority of observed surface stellar lithium (Li) abundances (Boesgaard \&
Tripicco 1986; Deliyannis et al.~1998; Deliyannis 2000).  The standard theory
does not include rotation, diffusion, mass loss, magnetic fields, etc.
Additional (non-standard) physical mechanisms apparently operate inside stars.
One of the goals of this study is to help elucidate this ``lithium problem''
(see Section 5.5).

The abundance of Fe has been determined in many clusters (e.g.,~Friel et
al.~2002).  Many open clusters have been studied for Li, beginning with Herbig
(1965) and more recently by Cummings et al.~(2012) and Francois et al.~(2013.
Abundances of many more elements in open clusters are important to develop a
more thorough picture of Galactic chemical evolution.  Studies of CNO, the
alpha elements, Fe-peak elements, and n-capture elements are critical.  For
this kind of research, both high spectral-resolution and high signal-to-noise
ratios (S/Ns) are needed.  Some recent work on this includes Maderak et
al.~(2013), Boesgaard et al.~(2013), Carrero \& Pacino (2011).

An especially fascinating group of open clusters is the super-metal-rich
trio NGC 6791, NGC 6253, and NGC 6583.  Studying their abundance patterns can
shed light on Galactic chemical evolution and potentially provide clues as to
how the Galaxy produced such high abundances.  These clusters also provide
invaluable opportunities to study Li depletion in super-metal-rich stars, and
to test theoretical predictions about the metallicity-dependence of Li
depletion, e.g.,~Deliyannis et al.~(1990), and the review by Pinsonneault
(1997).  In this study we focus on NGC 6791.

\section{NGC 6791}

NGC 6791 is a fascinating, unique cluster.  For an open cluster it is very
massive at $\sim$4000 $M_{\sun}$ (King et al.~2005).  Its age is $\sim$8.3 Gyr
with an uncertainty that is dominated by the uncertainty of the abundances
of C, N, and O (Brogaard et al.~2012).  The metallicity is at least twice that
of the sun (e.g.,~Boesgaard et al.~2009).  Its anomalies have lead to papers
that speculate that it is the ``Nucleus of a Tidally-Disrupted Galaxy?''
(Carraro et al.~2006), that it has ``Multiple Populations'' and even whether it
is an ``Open Cluster'' (Geisler et al.~2012), and that it has had ``Extended
Star Formation'' (Twarog et al.~2011).

The metallicity has been found to be super-solar with [Fe/H] $\sim$+0.4 from
spectroscopic studies of giant stars.  It is easier to observe the giant
stars because they are relatively bright compared to main-sequence and
turn-off stars.  First, Peterson \& Green (1998) found [Fe/H] = +0.4 from a
{\it bona fide} horizontal branch star.  Then, Worthey \& Jowett (2003) found
+0.32 from low resolution spectra ($R$ $\sim$2200) of 23 K giants.  Later,
Gratton et al.~(2006), Carraro et al.~(2006) and Carretta et al.~(2007) found
values of [Fe/H] of +0.39 to +0.47 from spectroscopic studies at higher
resolution.  Origlia et al.~(2006) found [Fe/H] = +0.35 in six M giants from
IR spectra with a resolution of 25,000.  Boesgaard et al.~(2009) were able to
determine abundances in two main-sequence turn-off stars from high-resolution
Keck/HIRES spectra ($R$ $\sim$45,000); they found [Fe/H] = +0.30 $\pm$0.02.

The photometry of NGC 6791 is excellent (e.g.,~Montgomery et al.~1994, Stetson
et al.~2003, and recalibrated photometry by Stetson reported in Brogaard et
al.~2012).  Ages from the H-R diagrams are 8-10 Gyr (Demarque et al.~1992,
Garnavich et al.~1994, Montgomery et al.~1994, Chaboyer et al.~1999, Brogaard
et al.~2012).  With differential reddening and data from eclipsing binary
stars, Brogaard et al.~(2012) refine the CMD and apply isochrones to determine
the age.  They find that the age is not sensitive to the precise value of
[Fe/H], but is sensitive to the uncertainty in [CNO/Fe].  Their best age
estimate is 8.3 $\pm$0.3 Gyr.  With its old age and high metallicity, it does
not conform to any age--metallicity relation for the Galactic disk.

The properties of NGC 6791 of super-high metallicity and old age are of
tremendous value in testing predictions from standard stellar evolution theory
(which does not include rotation, magnetic fields, diffusion, and mass loss)
regarding Li depletion in stars, see Cummings et al.~(2012).  One important
prediction is that cooler dwarfs should deplete more Li; this has been
verified in a substantial number of open clusters.  However, stars seem to
deplete a greater amount of Li than predicted, and do so during the
main-sequence phase of evolution (Jeffries \& James 1999; Deliyannis 2000;
Sestito \& Randich 2005; Cummings et al.~2012).  Various lines of evidence
suggest the extra Li depletion is due to rotationally induced mixing.  Some
examples: 1) short-period tidally locked binaries, Deliyannis et al.~(1994),
Ryan \& Deliyannis (1995); 2) the Li/Be and Be/B depletion correlations,
Deliyannis et al.~(1998), Boesgaard et al.~(2004), Boesgaard et al.~(2005); 3)
stars evolving from the turnoff to the giant branch, Sills \& Deliyannis
(2000).  Clearly there are other mechanisms can also play a role and must act
in addition to the effects of standard mechanisms.  Before we can effectively
decipher the relative contributions of non-standard mechanisms, it is
important to understand how well each of the standard predictions hold.

A key standard prediction is that dwarfs with higher metallicity should
deplete more Li (e.g.,~Deliyannis et al.~1990, Pinsonneault 1997).  Higher
metallicity leads to higher opacity near the base of the surface convection
zone (SCZ), which leads to a deeper SCZ, which then leads to more Li
depletion.  An additional prediction is that surface Li abundances should
dilute as stars evolve along the subgiant branch and the SCZ deepens past the
deepest point where Li has been preserved (Iben 1965, 1967).  This prediction
was investigated first by Herbig \& Wolff (1966).  It has been confirmed in
halo stars (Deliyannis et al.~1990, Ryan \& Deliyannis 1995), in globular
cluster stars (Lind et al.~2009), and open cluster stars (Anthony-Twarog et
al.~2013).  Also, some open cluster subgiants have provided insights about the
additional non-standard mechanisms operating inside of stars (e.g.,~M67, Sills
\& Deliyannis 2000).

NGC 6791 provides us with the unique opportunity to test some of these
predictions in the super-high-metallicity, old-age regime.  According to the
combination of these predictions, the best Li-preservers in this cluster
should be the {\it bluest} turnoff stars.  Cooler dwarfs are predicted to have
depleted more Li, and cooler evolving stars will have at least diluted their
Li and possibly depleted it even more, as is the case with M67.  Our goal was
thus to see how much Li remains in the bluest turnoff stars, if any, and then
to use any observed Li abundance pattern in further evolved stars to help
interpret which Li depletion mechanisms might be at work.

NGC 6791 has a heliocentric distance of $\sim$4 kpc (King et al.~2005).  It
has a Galactic latitude of +11${\arcdeg}$ which makes it 1 kpc above the
Galactic plane.  Most open clusters are near or in the Galactic plane.  Bedin
et al.~(2006) found the absolute proper motion of the cluster.  They determine
orbital parameters and find that it has a {\it boxy} orbit with high
eccentricity ($\epsilon$ $\sim$0.5.  Its perigalactic distance is about 3 kpc
and its apogalactic distance is about 10 kpc.  The orbital period is $\sim$130
Myr and it has crossed the Galactic plane several times.  It has remained as
an intact cluster due to its high mass and density.  It may have originated in
the inner regions of the Galaxy (Bedin et al.~2006).  Chemical evolution
models are able to produce [Fe/H] $\leq$ +0.30 for galactocentric distances of
$<$4-5 kpc even at early ages.  Following its formation in the inner part of
the Milky Way, NGC 6791 may have been ejected into its current high
eccentricity orbit by a massive feature, such as a bar (Jilkova et al.~2012).

NGC 6791 is in the Kepler-satellite field, thus Basu et al.~(2011) have
determined masses of evolved stars by asteroseismology.  They find 1.20
$M_{\sun}$ $\pm$0.01 for evolved stars.  Eclipsing binary systems have been
analyzed by Brogaard et al.~(2012) who find 1.15 $M_{\sun}$ $\pm$0.02 for
stars on the lower part of the red giant branch (RGB).

For this research we have obtained high resolution ($R$ = 46,000) spectra of
several more turnoff stars in addition to the two studied by Boesgaard et
al.~(2009) with the upgraded HIRES (Vogt et al.~1994) on Keck-1.  We report on
abundances of Fe and O, the alpha-elements Mg, Si, Ca, and Ti, the Fe-peak
elements Cr and Ni, and, as it turns out, upper limits on the abundances on
Li.

\section{OBSERVATIONS AND DATA REDUCTIONS}

Even though stars near the main-sequence turn-off are faint ($V\sim$17.4), we
chose to observe them rather than red giants in NGC 6791 for several reasons.
(1) The spectral features, especially Fe I, are weaker than in the giants.
(2) The lines are less blended and less saturated than in the super-metal-rich
low-gravity giants that have strong lines and crowded spectra.  (3) The
temperatures are warmer and similar to the Sun's temperature which makes
comparisons to the solar abundances more consistent.  (4) The atmospheres of
the turn-off stars are far less extended so the 1D analysis provides a better
approximation than it does in the giant stars.  (5) We wanted to establish
``baseline'' abundances for comparison with abundances in the evolved stars,
which may have dredged up products of nuclear processes to the surface.  (6)
We wanted to determine Li abundances as discussed in Section 2 above regarding
the Li issues.

In order to determine abundances for turn-off stars in NGC 6791, high spectral
resolution is advantageous; the super-solar metallicity means that many lines
are blended and/or saturated.  The turn-off stars have $V$ magnitudes near
17.4 which presents a challenge even for the Keck 10 m telescope and its
upgraded HIRES spectrograph.  Figure 1 shows the location of our sample in the
CMD.  They are all at or slightly evolved from the main-sequence turn-off.  We
will refer to them as turn-off stars to distinguish them from other
observations that have been made of red giant stars.  Our observations were
made on eight nights over three observing seasons.

Our observing strategy was to take multiple integrations of these faint stars
($V$ $\sim$17.4) and co-add them.  Each integration had to be short enough to
limit the number of cosmic-ray hits on the CCD, but long enough to get enough
signal to be able to add the spectra together matching the right wavelengths
in the line-crowded spectra.  Typical integrations were a total of 180 minutes
of 30 minute durations each.  For four stars that was spread over two nights.
We also tried to observe the NGC 6791 stars when they were near the meridian
at low airmasses in order to get the best S/N per integration.  The measured
spectral resolution per pixel was $\sim$46,000.  The typical S/N on individual
integrations was 16--17 and the total combined S/N $\sim$40.  The observing
log is in Table 1.  (Two of the 10 stars were obtained with the original
version of HIRES, analyzed by Boesgaard et al.~2009 and are reanalyzed here.)
The total S/Ns are per pixel and are determined from the co-added
spectra before they were fitted by a continuum.  They refer to the spectral
region near 6700 \AA\ and are typically $\sim$40.

On each of the four nights in 2008 and 2009 we took three spectra of Th-Ar
arcs, nine bias frames, nine quartz flat-fields at 3 s and nine quartz
flat-fields at 9 s.  The two sets of flats were needed to get proper exposures
for the three CCD chips.  The 3 s exposures were appropriate for the red and
green chips while 9 s exposures were needed for the blue chip, but were
saturated on the red and green chips.

The data from 2008 and 2009 were reduced using the HIRedux pipeline of
Prochaska\footnote{http://www.ucolick.org/$\sim$xavier/HIRedux} and various
IRAF\footnote{IRAF is distributed by the National Optical Astronomy
Observatory, which is operated by The Association of Universities for
Research in Astronomy, Inc., under cooperative agreement with the National
Science Foundation.} routines.  The pipeline was used to subtract the
master bias from the nine bias frames and to normalize the spectra with
our own images of the two master flat fields as well as spectrum extraction
and wavelength justification.  IRAF was used to co-add the multiple images and
to fit the continuum.

\section{STELLAR PARAMETERS}

We have used new photometry and data reductions to derive $T_{\rm eff}$ from
($B-V$).  The $B$ and $V$ values are from Stetson et al.~(2003) with a newer
unpublished reduction by Stetson as reported in Brogaard et al.~(2012).  The
overall reddening for NGC 6791 has been well-determined by Anthony-Twarog et
al.~(2007) who find $E(B-V)$ = 0.155 $\pm$0.016.  Differential reddening seems
to be present due to the fact that the spread in the sequences in the CMD
exceeds the photometric errors and is correlated with position on the sky
(Twarog et al.~2011; Platais et al.~2011; Brogaard et al.~2012).  The amount
of this differential reddening for NGC 6791 has been determined by Brogaard et
al.~(2012). (See Figure 2, in particular, in that paper.)  They find values for
the differential reddening along the line of sight to the cluster to be small,
within $\pm$0.04.  They do not directly determine overall reddening but adopt
a best estimate of $E(B-V)$ = 0.14 $\pm$0.02; this is consistent with the
reddening derived by Anthony-Twarog et al.~(2007).  Eight of our 10 stars have
determinations of the differential reddening.  We have used the calibrations
of Casagrande et al.~(2010) to determine temperatures; their calibration is
valid for dwarfs and subgiants through the full range of metallicities,
including super-metal rich stars.  Our values of $(B - V)_{0}$ include the
overall reddening of 0.155 from Anthony-Twarog et al.~(2007) and differential
reddening from Brogaard et al.~(2012).

As part of our Fe abundance analysis, we can also find temperatures
spectroscopically from the agreement of Fe lines with a range of excitation
potentials.  The mean difference between $T_{\rm eff}$s determined from the
colors and those from excitation potentials is $-$6 K $\pm$53 K.  Thus the
agreement between our photometrically based $T_{\rm eff}$ and the
spectroscopic-based ones is superb.  That standard deviation provides one way
to estimate the uncertainty in $T_{\rm eff}$.

We determined log $g$ from the basic relationship:

log $g$/$g$$_{\sun}$ = log $M/M$$_{\sun}$ + 4 log $T_{\rm eff}$/$T_{\rm
eff}$$\sun$ +0.4 ($M$$_{bol}$ $-$ $M$$_{bol}$${\sun}$)

For the solar values we used log $g$$_{\sun}$ = 4.438, $T_{\rm eff}$$\sun$ =
5778 K and $M$$_{bol}$${\sun}$ = 4.76.  The distance modulus for NGC 6791 is
13.51 $\pm$0.06 and the turn-off mass is 1.1 M$_{\sun}$ both from Brogaard et
al.~(2011).  The bolometric magnitude was calculated using the Stetson
$V$-band photometry, adjusted for the distance modulus of Brogaard et al.~plus
corrections from Table 2 of Masana et al.~(2006) for FGK stars.  Individual
corrections were obtained by interpolating between the [Fe/H] = 0.00 and
[Fe/H] = +0.50 columns for our [Fe/H] = +0.30, and the photometrically
determined $T_{\rm eff}$ of each star.

Values for the microturbulent velocity, $\xi$, were found from the empirically
determined relationship by Edvardsson et al.~(1993) from 189 F and G disk
dwarfs with dependencies on both $T_{\rm eff}$ and log $g$.  The colors and
stellar parameters are given in Table 2.  We made preliminary model
atmospheres with those parameters and [Fe/H] = +0.35.  We refined that value
based on our measurements of Fe I and Fe II which gave [Fe/H] = 0.30 $\pm$0.04
(see Section 5 for those details).  We used model atmospheres of Kurucz
(1993).

\section{ABUNDANCES AND RESULTS}

\subsection{Iron}

We measured the equivalent widths for Fe I and Fe II lines which were
unblended and unsaturated. The lines are unsaturated when log $W$$/$$\lambda$
$\leq$ $-$4.82.  For the eight stars with observations with the upgraded
version of HIRES we measured 37-46 Fe I lines and 5-8 Fe II lines; for the two
stars from the earlier version of HIRES we measured 26 lines of Fe I and 6 of
Fe II.  Figure 2 shows two spectral regions of MJP 303 with three lines of Fe
I in each region.  The spectra have S/Ns of $\sim$40 and the Fe I lines
are clear, unblended, and unsaturated.  Due to severe line-crowding (and lower
signal), we did not measure any Fe lines on the blue chip ($\sim$4100--5240
\AA).  The wavelength interval we used for Fe I lines was 5560--7912 and for
Fe II was 5534--7711 \AA.  The Fe I and Fe II lines and their excitation
potentials and log $gf$ values are given in Table 3.  The Table also gives the
measured equivalent widths for Fe I and Fe II lines in two of the stars:
MJP 303 and MJP 2279.

We used the {\it abfind} driver in the updated version of
MOOG\footnote{http://www.as.utexas.edu/~chris/moog.html} (Sneden 1973) to
determine abundances of both Fe I and Fe II.  Table 4 shows the [FeI/H]
results for each star from Fe with the errors from the agreement among the
lines in the sample.  The [FeI/H] agrees very well for 8 of the 10 stars,
but the other two, MJP 5597 and MJP 6930, were significantly lower.  As
can be seen from Table 4 the [FeI/H] abundance for MJP 5597 is $-$0.05
$\pm$0.11, or over 3$\sigma$ below the cluster mean and for MJP 6930 is
$-$0.33 $\pm$0.11 nearly 6$\sigma$ below the cluster mean.  Those two
stars appeared to be turn-off stars based on their location in the
CMD and position within the cluster, but their
[FeI/H] abundances are discrepant.

We also examined the radial velocities for the 10 stars in our sample.  The
cluster radial velocity was found to be $-$48 $\pm$9 km s$^{-1}$ by Friel et
al.~(1989).  Bedin et al.~(2006) found $-$47.1 $\pm$0.7 km s$^{-1}$.
Recently, Gao \& Chen (2012) have determined the cluster radial velocity from
the 95 stars with high membership probabilities and found $-$46.4 $\pm$0.2 km
s$^{-1}$.  The radial velocities for those two stars were found $-$60 km
s$^{-1}$ for MJP 5597 and +5 km s$^{-1}$ for MJP 6930.  The radial velocities
that we measured from our spectra of the other eight stars range from $-$45 to
$-$53 with a mean of $-$47.6 $\pm$2.8 km s$^{-1}$ in agreement with the
cluster mean.  The two stars, MJP 5597 and MJP 6930, are considered
non-members based on their radial velocities and their Fe abundances.  We have
excluded them from the rest of the analysis.

We determine the mean [Fe/H] for the eight member stars from both Fe I and Fe
II.  We found the stellar mean by weighing Fe abundance by the number of lines
measured in each ionization stage.  The stellar mean from Fe I is +0.30
$\pm$0.04 and from Fe II is +0.30 $\pm$0.05.  The agreement between the Fe
abundance from the two ionization states is excellent with a mean difference
of 0.00 $\pm$0.04. The final mean for eight stars is [Fe/H] = +0.30 with the
standard deviation of the mean of $\pm$0.02.

Previously, abundances of Fe have been determined primarily from evolved stars
due to their relative brightness.  Peterson \& Green (1998) analyzed the
spectrum of one horizontal branch star at a resolution of 20,000 and found
[Fe/H] = +0.4 $\pm$0.1.  Worthey \& Jowett (2003) observed 24 giants, but at a
resolution of only $R$ = 2200, and found [Fe/H] = +0.32 $\pm$0.02.  From
higher-resolution spectra ($R$ = 29,000) of four red giants Gratton et
al.~(2006) found [Fe/H] = +0.47 $\pm$0.08 and Carretta et al.~(2007) found
[Fe/H] = +0.47 $\pm$0.07 from the spectra of the same four giants.  Carraro et
al.~(2006) analyzed 10 K giant stars at the top of the RGB with WIYN/Hydra
spectra at 17,000 resolution and found [Fe/H] = +0.38 $\pm$0.02.  The infrared
spectra of six M giants taken with NIRSPEC of Keck-2 at a resolution of $R$ =
25,000 were analyzed by Origlia et al.(2006) who found [Fe/H] = +0.35
$\pm$0.02.  More recently Geisler et al.~(2012) derived [Fe/H] = +0.42
$\pm$0.01 from an array of stars on the RGB (including some red clump stars)
with spectra from WIYN/Hydra and Keck/HIRES.  Gao \& Chen (2012) found [Fe/H]
= +0.32 $\pm$0.11 from SDSS-DR8 spectra of 87 stars at $R$ = 2200 resolution.
Boesgaard et al.~(2009) obtained spectra with Keck/HIRES at $R$ $\sim$45,000
resolution of two turn-off stars and derived [Fe/H] = +0.30 $\pm$0.01 from Fe
I and Fe II lines.

We consider our value of [Fe/H] = +0.30 $\pm$0.02 to be the best [Fe/H]
abundance for NGC 6791 determined to date because it is based on unblended and
unsaturated lines of both Fe I and Fe II in eight main-sequence turn-off stars
observed at a resolution of $R$ = 46,000 and S/N of $\sim$40.  The atmospheres
of turn-off stars can be more directly compared to the sun than the extended
atmospheres of the giant stars.

\subsection{Oxygen}

The O I triplet lines at 7774 \AA\ were present on the red chip spectra of the
stars taken with the upgraded HIRES, but not on the spectra with the settings
we used on the earlier HIRES version.  Another high excitation O I line
occurs at 6158 \AA, but that line was too weak to measure in any of the eight
stars.  We also tried to synthesize the [O I] line at 6300 \AA\ but found it
too weak to give reliable O abundances.  Furthermore, with our grating
settings that line is very close to the edge of the CCD chip where continuum 
fitting can be a problem.

In order to make a differential analysis with respect to the Sun, we used a 10
s exposure of the Moon that we had taken with Keck/HIRES in 2003 January 11
UT.  The measured S/N in the region of the O I triplet is 350.  Figure 3 shows
an 11 \AA\ segment of this spectral region of the solar/lunar spectrum and
that of two of the stars in NGC 6791.  We did a spectrum synthesis on the O I
triplet lines, but found that a single O abundance does not fit the three
lines.  This is because the non-LTE corrections are different for each of the
three lines.  We measured the equivalent widths for the 3 lines in the lunar
and stellar spectra and derived the LTE abundance for each line.  We used log
$gf$ values from Wiese et al.~(1996) of 0.369, 0.223, and 0.002 for the three
lines, 7771.9, 7774.2, and 7775.4.  For the Sun, we measured 62.8, 59.8, and
46.6 m\AA\, respectively, and derived an LTE abundance of log N(O) = 8.74
$\pm$0.05 with the standard solar model of 5770 K, 4.44, 0.00, 1.0 km
s$^{-1}$.  This is in good agreement with the solar O abundance derived by
Ramirez et al.~(2013) of 8.77 $\pm$0.01.

Takeda (2003) has determined the nLTE corrections for each O I line for an
array of temperatures, log g values, and Fe abundances.  We applied the
appropriate corrections for each line by interpolating between Takeda's tables
for our parameters.  We note that our measured [Fe/H] of +0.30 is above the
high end of Takeda's metallicity at +0.25, thus we used that value.  However
he nLTE correction is not very sensitive to [Fe/H].  For Takeda's sample of
$\sim$300 stars the nLTE correction for [Fe/H] varies by less than 0.3 dex
over the entire range of their sample's metallicities from about $-$3.0 to
about +0.5; this can be seen in Figure 8(b) of Takeda (2003).  We then found a
mean abundance for O from the three lines.  The nLTE corrections make the LTE
O abundances lower.  We applied his nLTE corrections for the Sun and derived
log N(O) = 8.65 $\pm$0.07.  This is in excellent agreement with the nLTE solar
value derived by Ramirez et al.~(2013) of log N(O) = 8.64 and by Asplund et
al.~(2009) of 8.69 $\pm$0.05.

We show the results in Table 5.  For each star, we give the equivalent width of
the three lines in the O I triplet, the value for [O/H] for each line, the
correction for the effects of nLTE as calculated from Takeda (2003), the value
for the corrected O abundance as [O/H]$_{\rm n}$, and the final [O/Fe]$_{\rm
n}$.  (For MJP 1328, we could only reliably measure the line at 7774 \AA\ and,
for MJP 1346, we could not determine the strength of the line at 7775 \AA.)
The corrections to the LTE values of [O/Fe] range from $-$0.04 to $-$0.09 with
a mean correction of $-$0.06 $\pm$0.02.  We give the mean and standard
deviation of [O/Fe]$_{\rm n}$ for each star (except MJP 1328 with only one
line measurable).  Our values are somewhat below solar with a mean of
[O/Fe]$_{\rm n}$ = $-$0.06 $\pm$0.02.

The results from the individual O I lines are shown graphically in Figure
4.  This shows that no one line is systematically lower/higher than
the others in the six stars and that no one star is deviant.  Also plotted are
the standard deviations in [O/Fe]$_{\rm n}$ and the standard deviation of the
mean for each of the three O I lines in each star.

In Figure 5 we show a comparison of our [O/Fe]$_{\rm n}$ versus[Fe/H] with
results for field stars.  The main field star sample is from Ramirez et
al.~(2013), selected to have [Fe/H] greater than 0.0 and ages older than 7.7
Gyr.  Those 34 stars are from their total sample of 835 nearby FGK stars.
These comparison stars and their abundances are given in Table 6.  They
also determined O abundances from the O I triplet and applied nLTE corrections
to each line.  Also shown in the figure are the [O/Fe] and [Fe/H] from the
sample of old, metal-rich stars of Chen et al.~(2003).  These field stars are
between 8--10 Gyr and have [Fe/H] values $>$ 0.00.  Although Chen et al.~did
not correct for nLTE effects, they did do a differential analysis with respect
to the Sun.  As we have shown above, those corrections are small, about
$-$0.06 dex.

It can be seen in this figure that our [O/Fe]$_{\rm n}$ values in the
turn-off stars in NGC 6791 are generally below the majority of the counterpart
field stars.  It also indicates that NGC 6791 turn-off stars fit the pattern
of decreasing [O/Fe] with increasing [Fe/H], similar to the results of Chen et
al.~(2003).  Both Figures 4 and 5 show that there is no sign of an intrinsic
spread in [O/Fe]$_{\rm n}$ in our six turn-off stars, which is consistent with
a single population.

The Galactic trend of [O/Fe] with [Fe/H] is high at low metallicities: [O/Fe]
$\sim$+1.0 at [Fe/H] of $-$3.5 and then declines to near zero at [Fe/H] = 0.0
(e.g.,~Boesgaard et al.~2011).  The results in Figure 5 appear to continue that
trend to higher values of [Fe/H].  It is thought that O is formed via massive
stars and SN II, which are dominant in the early days of the Galaxy, and those
stars have low Fe/H.  The Fe is formed primarily in intermediate mass stars
and SN Ia.  Jilkova et al.~(2012) suggest that NGC 6791 could have originated
in the central region of the Galactic disk where there would be more rapid
metal-enrichment subsequently migrated out to its current orbit.

Gratton et al.~(2006) analyzed four red clump stars in NGC 6791 with V
magnitudes of $\sim$14.6.  Their spectral resolution was $R$ = 29,000 and S/N
$\sim$60.  They were able to use the [O I] line at $\lambda$6300, which is
much stronger in giant stars than in dwarf or turn-off stars.  They found
[Fe/H] = +0.47 and [O/Fe] = $-$0.32.  Carretta et al.~(2007) analyzed the same
four giants and found [O/Fe] = $-$0.31 $\pm$0.08, with the mean [Fe/H] at
+0.47.  

Abundances of O were also found by Geisler et al.~(2012) from spectrum
synthesis of the [O I] line.  They found a range in [O/Fe] from $-$0.10 to
+0.25 in their red giant stars.  Origlia et al.~(2006) used the IR features of
OH near 1.6 $\mu$m from spectra of six M giants taken at a spectral resolution
of 25,000 with NIRSPEC on the Keck-2 telescope to find [O/H]; with their
values of [Fe/H] = +0.35 $\pm$0.02 they find [O/Fe] = $-$0.07 $\pm$0.03.
These two [O/Fe] results seem to be in conflict with the results of Gratton et
al.~(2006) and Carretta et al.~(2007) for giant stars.

Geisler et al.~(2012) found evidence for two stellar populations in NGC 6791
based on eight stars with low [Na/Fe] (mean = $-$0.15) and 13 with high
[Na/Fe] (mean = +0.38).  All of the five stars they measured at the base of
the RGB are in the low-Na group.  In the low-Na stars the range
in [O/Fe] is from $-$0.10 to +0.06.  The high-Na group has a spread in [O/Fe]
of $-$0.10 to +0.25; the two groups are distinct in [Na/Fe] but there is
overlap in [O/Fe].  Their range in [O/Fe] covers +0.35 dex.

The question of multiple populations in NGC 6791 has recently been examined by
Bragaglia et al.~(2014).  They found no clear anti-correlation between Na and
O.  They used our spectra and accepted the automatic data reduction done by
the archivists.  (See our description of our data collection strategy and our
reduction in Section 3, especially our use of the two different master flat
fields of 3 s and 9 s as appropriate for the three CCDs of HIRES.)  They tried
to synthesize the [O I] line at 6300 \AA\ to find the O abundance.  At the
beginning of this section we argue against using that line due to its weakness
in turn-off stars and the fact that it is too close to the edge of the CCD to
be reliable.  A firm knowledge of the strength of the CN blends in the region
of the [O I] lines is also necessary.  They determined O from [O I] in four of
our stars for which we have O from the O I triplet.  The four stars are MJP
303, 885, 1328 and 2279 with corresponding SBG numbers 15592, 14416, 13334,
11220.  Our values for [O/Fe]$_{\rm n}$ are $-$0.10, $-$0.10, 0.00, and
$-$0.07, respectively for an average of $-$0.068 $\pm$0.027.  Their values
show a wide range with [O/Fe] values of $-$0.05, $-$0.26, $-$0.08 and +0.08,
respectively averaging $-$0.078 $\pm$0.082.

For those four stars our [O/Fe] from the O I triplet has a range of 0.10 dex
while theirs, from [O I] has a range of 0.34 dex.  There are seven red giant
stars in their sample for which they find O abundances from HIRES spectra
using [O I].  The values for [O/Fe] in those seven stars range from +0.10 to
$-$0.26 with a mean of $-$0.11 $\pm$0.11.

Recently Cunha et al.~(2015) report on O and Na abundances in giants in
NGC 6791 from the IR spectra of 11 red giants and red clump stars determined
from the vibration-rotation features of OH.  This requires knowledge of the C
and N abundances as well.  They find Na abundances from two lines of Na I in
the $H$ band.  They do not find the Na-O anticorrelation found by Geisler et
al.~(2012) and find a solar ratio for [O/Fe].

\subsection{Alpha-Elements}

We measured the equivalent widths of 7-9 Si I lines, 7-9 Ca I lines and 6-8 Ti
I lines in all eight stars.  For Mg I there were only 2-3 lines for the stars
observed with the upgraded HIRES and none on the spectra of the other two
stars.  (The Mg I lines at $\lambda$6965 and $\lambda$7387 were too far to the
red to be covered with the settings we used and the line at $\lambda$6318 fell
in a gap between the echelle orders.)  The cluster mean values are [Mg/Fe] =
+0.08 $\pm$0.04, [Si/Fe] = +0.07 $\pm$0.02, [Ca/Fe] = $-$0.13 $\pm$0.04, and
[Ti/Fe] = $-$0.04 $\pm$0.03.  The detailed information is given in Table 8
along with the results for [Fe/H] and [O/Fe]n.  In that table we give the
sample standard deviation, $\sigma$, and the standard deviation of the mean,
$\sigma$$_{\mu}$.

Figure 6 shows a plot of each of the four alpha-elements relative to Fe in
each star as compared to the old and metal-rich field stars.  The comparison
samples are from: (1) Chen et al.~(2003) who studied nine old, metal-rich
stars in the solar neighborhood, (2) Edvardsson et al.~(1993), who analyzed
189 F and G dwarfs from which we culled six that are old and metal-rich, (3)
Feltzing \& Gonzalez (2001) who studied eight super-metal-rich stars of which
four are older than 8 Gyr, and (4) Reddy et al.~(2006), who determined
abundances of 22 elements in 176 nearby F and G stars from which we found six
that had ages $>$7.7 Gyr and [Fe/H] $\geq$0.00 as determined by Ramirez et
al.~(2012).  The comparison stars and abundances are given in Table 7.  There
is considerable spread in the field star values.  In both the cluster and
field star samples, the ratios of [Mg/Fe] and [Si/Fe] are enhanced at
$\sim$+0.05, while [Ti/Fe] is approximately solar in the mean.  Two of our
eight stars, 885 and 1783, seem low in [Ca/Fe], while the others match the
field star distribution at solar and somewhat below solar.  The two stars are
lower than the mean of all eight stars by $\sim$1 sigma and lower than the
mean of the other six stars by $\sim$2 sigma.

The abundances of some of the alpha-elements in giant stars in NGC 6791 have
been determined by Carraro et al.~(2006) and Carretta et al.~(2007).  The
results for [Si/Fe], [Ca/Fe], and [Ti/Fe] in the 10 giants of Carraro et
al.~is in agreement with those in our eight turn-off stars within the errors
(see Table 9).  Four red clump giants were analyzed by Carretta et al.~for all
four alpha elements.  Those results are in agreement with ours and they
also find a low value for [Ca/Fe] of $-$0.15 $\pm$0.08.  Origlia et
al.~(2006) find the alpha-element abundances relative to their value of [Fe/H]
of +0.35 $\pm$0.02 to be near solar.  We conclude that there are no observable
differences in the alpha-elements between the turn-off and the giant stars
that are due to the effects of nuclear burning followed by dredge-up in the
giants.

\subsection{Fe-Peak Elements}

We were able to measure 5-6 lines of Cr I.  They were unsaturated lines in all
eight stars.  For Ni I we could measure between 11 and 16 lines, all of which
were unsaturated.  (The four lines longward of 7000 \AA\ were unavailable on
MJP 4112 and 5061.) The values found for [Cr/Fe] and [Ni/Fe] are given in
Table 8.  It can be seen that both Fe-peak elements are similar to Fe in our
turn-off stars.

We have compared our values of [Cr/Fe] and [Ni/Fe] with samples of field stars
that are old and metal-rich.  This can be seen in Figure 7.  This comparison
hints that the NGC 6791 stars might possibly be slightly enriched in Cr and Ni
with the mean for [Cr/Fe] = +0.05 $\pm$0.02 and [Ni/Fe] = +0.04 $\pm$0.01.
Our value for [Ni/Fe] is not significantly different from that found in the
giant stars as given in Table 9.

\subsection{Lithium}

We used the {\it synth} driver in MOOG to determine Li abundances by spectrum
synthesis.  Examples of the synthesized spectra are shown in Figure 8.  There
were no detections of Li in any of these rather cool turn-off stars.  The
upper limits are given in Table 8.  Their position in the Li-temperature plane
is shown in Figure 9.  These are compared with the Li detections and upper
limits in turn-off stars in NGC 6253, also a rather old, super-metal-rich
cluster.  The Li abundances are given as A(Li) = log N(Li)/N(H) + 12.00.

For our comparison cluster, NGC 6253, Bragaglia et al.~(1997) found an age of
$\sim$3 Gyr; Piatti et al.~(1998) determined an age of 5 $\pm$1.5 Gyr; Sagar
et al.~(2001) obtained an age of 2.5 $\pm$0.6 Gyr.  Using isochrones with
alpha-element enhancements, Twarog et al.~(2003) found an age of 3 $\pm$0.5
Gyr as did Anthony-Twarog et al.~(2010).  This cluster is less than half the
age of NGC 6791.  However, it may be even more metal-rich.  Carretta et
al.~(2007) found [Fe/H] = +0.46 $\pm$0.03 from four red clump giants while
Sestito et al.~(2007) derived +0.36 $\pm$0.07.  The Fe abundance from 38
turnoff members was found to be [Fe/H] = +0.43 $\pm$0.01 by Anthony-Twarog et
al.~(2010) from Hydra spectra at the WIYN telescope; they also analyzed 18
giant stars and found [Fe/H] = +0.46.  NGC 6253 and NGC 6791 are two clusters
with reliable [Fe/H] determinations in the range +0.3 - +0.5 that have been
studied for Li.  Thus NGC 6253 makes a good comparison cluster for NGC 6791.
Abundances and upper limits of Li were determined by Cummings et al.~(2012).
There is a third cluster that is super-metal-rich: NGC 6583.
Unlike NGC 6253 and 6791, whose super-metal-richness is evidenced by
multiple studies, only two stars in a single study suggest that NGC 6583 is
super-metal-rich (Magrini et al.~2010); furthermore, no Li abundances have
yet been reported for this cluster.  However, if its status as a
super-metal-rich cluster becomes more robust from additional studies, its
younger age of ~1 Gyr (Carraro et al.~2005) would make it a very interesting
target for Li studies.

Figure 10 shows that the Li abundances and upper limits show similar patterns
in the two old, metal-rich clusters.  Both clusters show Li upper limits in
the $T_{\rm eff}$ range 5800--5400 K, with values of A(Li) are $\leq$ 1--1.5
that are well below the presumed initial values of A(Li) $>$3 and the
detections seen in hotter subgiants of NGC 6253.  In NGC 6791, stars hotter
than 5800 K evolved off the main sequence long ago, and, unfortunately, there
is no longer any record of what kind of Li abundances such stars may have
contained.  It should be noted that NGC 6253's younger stars in the $T_{\rm
eff}$ range 5800--5400 K have masses near 1.55 $M$$_{\sun}$ (Figure 5 of
Cummings et al. 2012), whereas the stars in NGC 6791 have lower masses near
1.15 $M_{\sun}$ (Brogaard et al.~(2012).  Thus, although both sets of stars
show similar upper limits today, they originate from different regions of the
main sequence A(Li) versus $T_{\rm eff}$ (or A(Li) versus mass) diagram, and
may have experienced different Li depletion histories during the main
sequence.

We can compare the Li in these two clusters as a function of stellar mass, as
was done by Cummings et al.~(2012) in their Figures 4(a) to 8(a).  This is
shown for NGC 6791 and NGC 6253 in Figure 10.  The features shown for NGC 6253
have been attributed by Cummings et al.~to the following. (1) Stars with
$M_{\sun}$ of 1.5 to 1.55 are evolving from the cool side of the Li-dip found
in main-sequence stars.  They are the most evolved stars, and show no
detectable Li.  (2) Stars with masses between 1.35 and 1.5 $M_{\sun}$ are still
on the main sequence and show a bimodal Li distribution.  (3) A "Li plateau"
(Deliyannis 2000) is found in main-sequence stars cooler than the Li dip
itself, yet warmer than the onset of Li depletion in G dwarfs that increases
with decreasing mass.  These stars have been able to preserve the most Li.  (4)
Stars with masses less than 1.2 $M_{\sun}$ show the effects of main-sequence
depletion with the lowest mass stars having the greatest depletion.

The stars in that mass range ($<$1.2 $M_{\sun}$) in the older cluster, NGC
6791, have evolved off the main sequence.  These stars depleted their Li
during main-sequence phase while they were G-dwarfs and, therefore, we can only
determine upper limits on A(Li).  That G-dwarf Li-decline has been observed in
many open clusters, e.g.~Cayrel et al.~(1984) in the Hyades.  We use the mass
for these stars as determined by Brogaard et al.~(2012) at 1.15 $M_{\sun}$.
(We have found that same mass with the Yale--Yonsei isochrones assuming [Fe/H]
= +0.30, [alpha/Fe] = 0.0, and the turnoff $T_{\rm eff}$ of 5750 K to be
consistent with the mass determinations of Cummings et al.~(2012).)  Our NGC
6791 stars have only recently evolved off the main sequence and they are
similar to the Li-mass trend of NGC 6253.

Both the age and the metallicity of a cluster are known to affect the amount
of Li depletion in main-sequence stars (e.g.~Herbig 1965, Balachandran 1995).
The older the cluster, the longer the time it has had to undergo the
``slow-mixing'' that depletes Li.  The more metal-rich the cluster is, the more
Li depletion it will have undergone according to theory (Deliyannis et
al.~1990; Pinsonneault 1997).  Both of these influences seem to have been
working on NGC 6791 at 8.3 Gyr and [Fe/H] = +0.30.  Figure 5(a) of Cummings et
al.~(2012) also shows stars of the solar-metallicity cluster M67, and it is
quite clear that our upper limits of NGC 6791 are well below the A(Li)
$\sim$2.6 exhibited by M = 1.15 $M_{\sun}$ dwarfs in M67, consistent with the
predictions of standard theory.  However, M67 is only half as old as NGC
6791, so perhaps the extra Li depletion in NGC 6791 is due to its older age.
On the other hand, M67 and NGC 6253 have similar ages, but the more
metal-rich NGC 6253 does have less Li than M67 at 1.15 $M_{\sun}$.  We can
bring to bear the values of A(Li) found in NGC 188 which is of roughly
solar-metallicity and with an age of 7-8 Gyr. There are some Li detections at
the turnoff for NGC 188 (Hobbs \& Pilachowski 1988; Randich et al.~2003).
These turnoff stars in NGC 188 have $T_{\rm eff}$ = 5900--5750 K, comparable to
that of NGC 6791, but of slightly lower masses of M = 1.0-1.1 $M_{\sun}$, due
to NGC 188's lower metallicity.  Regardless, the NGC 188 turnoff stars have
A(Li) = 2.0--2.6, which is clearly higher than our upper limits in NGC 6791,
once again consistent with the general expectations from standard theory that
higher metallicity stars have greater Li depletion.

We speculate on the spectacular possibility that NGC 6791 might be the {\it
only} star cluster in which {\it not even one star} has been able to preserve
its initial lithium content.  Dwarf stars in NGC 6791 would be even cooler
than the warmest stars in our sample, originating from the even further
Li-depleted region of the steep Li-mass trend, and thus have undetectable Li.
Plus, more evolved subgiants and giants would have experienced additional Li
depletion due to dilution and mixing past the RGB clump, as seen in other
clusters.  Any future detection of Li in NGC 6791 would thus be both
surprising and interesting.  To have no surviving Li in any member star of a
cluster seems to require both very high metallicity and very old age, and may
thus be seen again in the future if, in the possibly unlikely event,
other very old super-metal-rich clusters are discovered.

\subsection{Error Estimates}

The values given for $\sigma$ in Table 4 represent the standard deviation of
each Fe I or Fe II line from the mean [FeI/H] or [FeII/H] for each star.  We
have made determinations of the errors due to uncertainties in the stellar
parameters.  Our models change one parameter at a time with the following
uncertainties: $\pm$75 K in $T_{\rm eff}$, $\pm$0.20 in log $g$, $\pm$0.10 in
[Fe/H], and $\pm$0.20 km s$^{-1}$ in the microturbulent velocity.  These are
given for two representative stars in Table 10.  The two stars, MJP 303 and MJP
885, encompass the range in parameters.  (The one cooler star with lower log
$g$ is MJP 5061 for which our data come from the earlier version of HIRES and
do not include the full range in wavelength, nor do we have abundances of O or
Mg in this star.)

There is almost no dependence of the final abundances on the value of [Fe/H].
Only Fe II and O I have much dependence on log $g$.  The increase in $T_{\rm
eff}$ of 75 K leads to a decrease in abundance for all but Fe II and O I of a
few hundredths of a dex.  The increase of +0.20 km s$^{-1}$ increases all of
the elemental abundances by +0.02 to +0.07.  The square root of the sum of the
squares of the individual errors is given in the final column.  Inasmuch as
the errors are not independent of each other, the number given as the total
error is an overestimate.

\section{SUMMARY AND CONCLUSIONS}

We have made high-resolution observations with the Keck 10 m telescope and
HIRES of turn-off stars in the old, metal-rich open cluster, NGC 6791, to
determine abundances of several elements.  We have selected spectral features
that were unblended even though there is much line crowding in the metal-rich
stars.  Our S/N values of $\sim$40 and the resolution of $R$ = 46,000 helped
to insure clean features.  All of the lines we measured were longward of 5500
\AA.

The spectra of our stars at the main-sequence turn-off have far less blending
than those of the giants that have been used in previous determinations of Fe
abundance.  We derived Fe abundances from both Fe I and Fe II
using only unsaturated and unblended lines.  There were 26-42 lines of Fe I
and 5-8 lines of Fe II in eight turn-off stars.  Both Fe I and Fe II gave the
same Fe abundance, [Fe/H] = +0.30 $\pm$0.02.  The quality of the spectra, the
observations made of turn-off stars, the number of stars observed, the use of
a large number of unblended and unsaturated Fe I and II lines combine to make
this the best determination of [Fe/H] in NGC 6791 to date.

To find the abundance of O we have used the O I triplet at 7774 \AA.  We have
used our spectrum of the Moon from Keck/HIRES as a surrogate for the Sun to
determine the solar O abundance, both in LTE and in nLTE.  Those respective
values are 8.74 and 8.65.  Then we found [O/H], [O/Fe], and [O/H]n and [O/Fe]n
for six turn-off stars.  The nLTE value for [O/Fe]$_{\rm n}$ is $-$0.06
$\pm$0.02.  This value, slightly lower than solar, makes NGC 6791 fit well
with the trend of decreasing O with increasing Fe.

We have compared our results for the alpha elements, Mg, Si, Ca, and Ti, with
those in a sample of old and metal-rich field stars and found that they are
all near solar, within $\pm$0.1 dex in contrast to our O results.  The Fe-peak
elements are also near solar with [Cr/Fe] = +0.05 $\pm$0.02 and [Ni/Fe] = 0.04
$\pm$0.01.  We compared our results for seven elements with two analyses of
giant stars and found no evidence of dredged-up products of nuclear fusion in
the giants.  Our abundances of these seven elements provide benchmark
abundances for this cluster.

None of the turn-off stars showed a Li I line thus we could determine only
upper limits on A(Li); they were between $<$0.9 to $<1.7$.  We have compared
our Li results with those of NGC 6253, another relatively old,
super-metal-rich cluster as a function of both $T_{\rm eff}$ and stellar mass.
We have demonstrated that both old age and high metallicity contribute to the
Li depletion in the 1.15 $M_{\sun}$ turn-off stars.  It is probable that all
of the stars in NGC 6791 have lost whatever Li they had initially.

\acknowledgements We thank Gabriel Dima and Jeffery A. Rich for their help
with the data reduction.  We were aided in the data collection by Jeffery
A.~Rich and Brendan P.~Bowler.  This research was supported by NSF grant
AST-1211699 to C.P.D.

\begin{figure}
\plotone{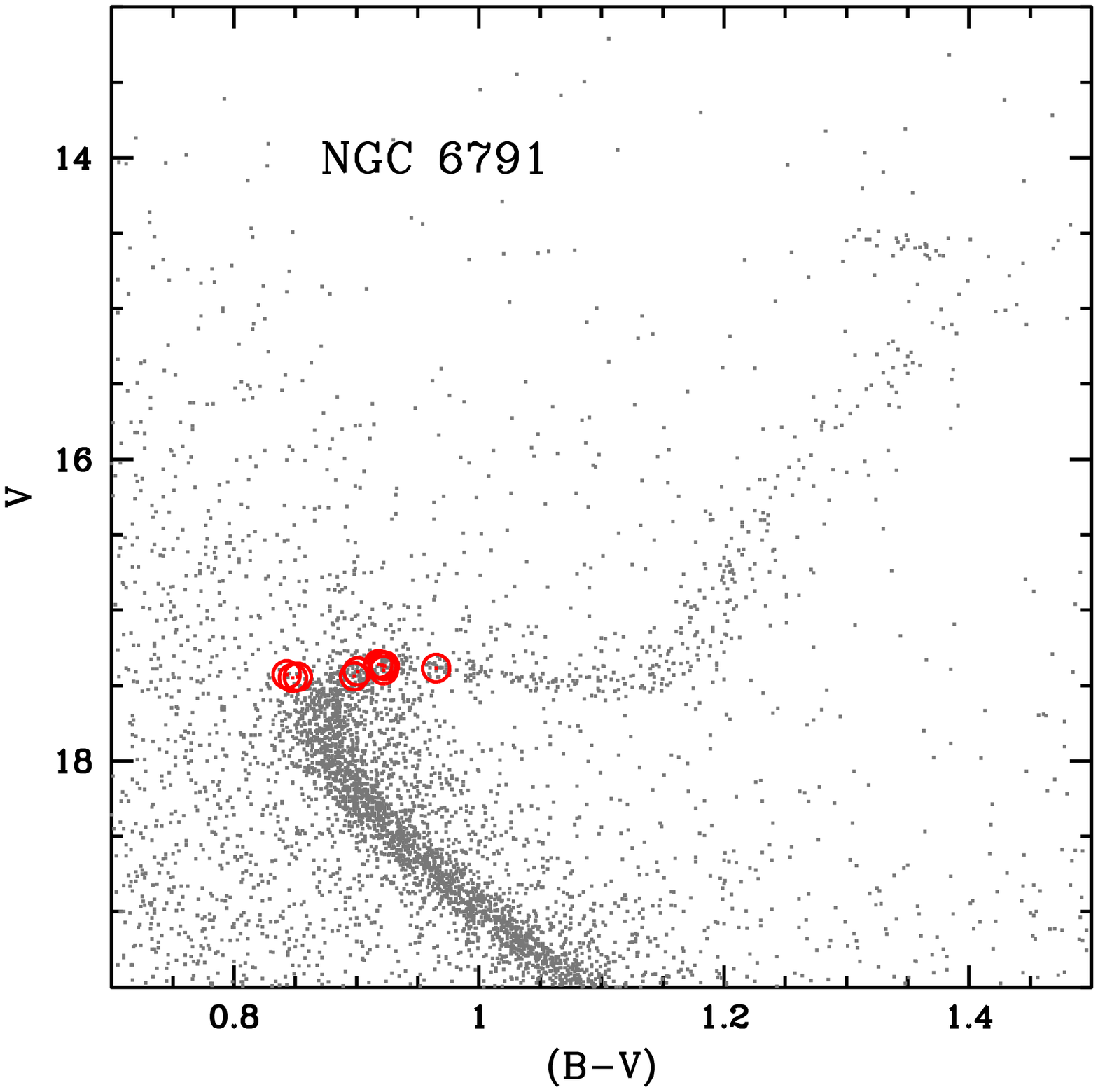}
\caption{Color--magnitude diagram for NGC 6791 highlighting the stars we
observed.  The photometry is from Stetson as recalibrated and reported in
Brogaard et al.~(2012).}
\end{figure}

\begin{figure}
\plotone{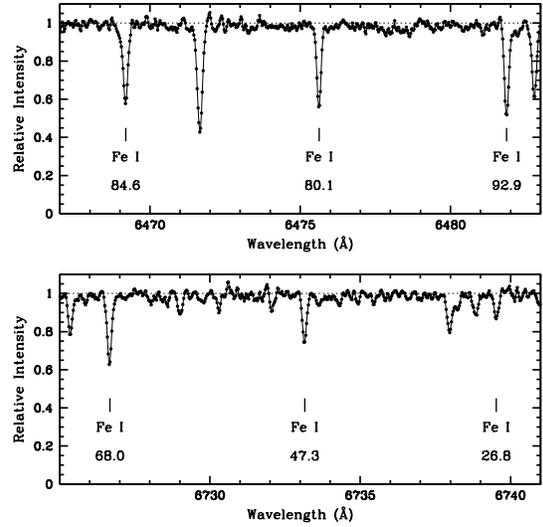}
\caption{Examples of unblended, unsaturated Fe I lines in MJP 303 in two
spectral regions.  The measured equivalent widths are given below each line in
m\AA.  The continuum level is shown by the dotted line at 1.0.}
\end{figure}

\begin{figure}
\plotone{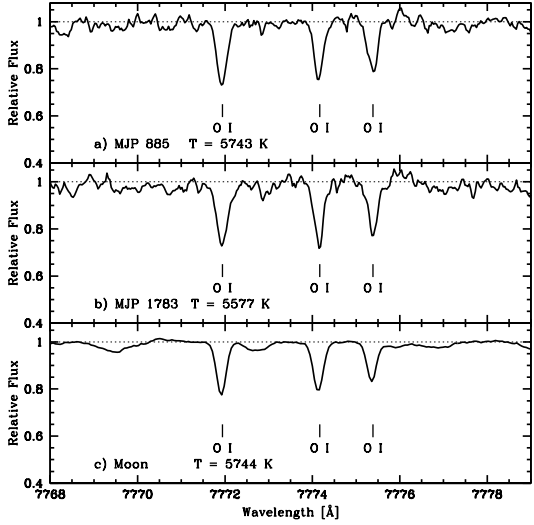}
\caption{ An 11 \AA\ region of spectra in the region of the O I triplet in
two stars in NGC 6791 and in our lunar/solar spectrum that was also taken with
Keck/HIRES.}
\end{figure}

\begin{figure}
\plotone{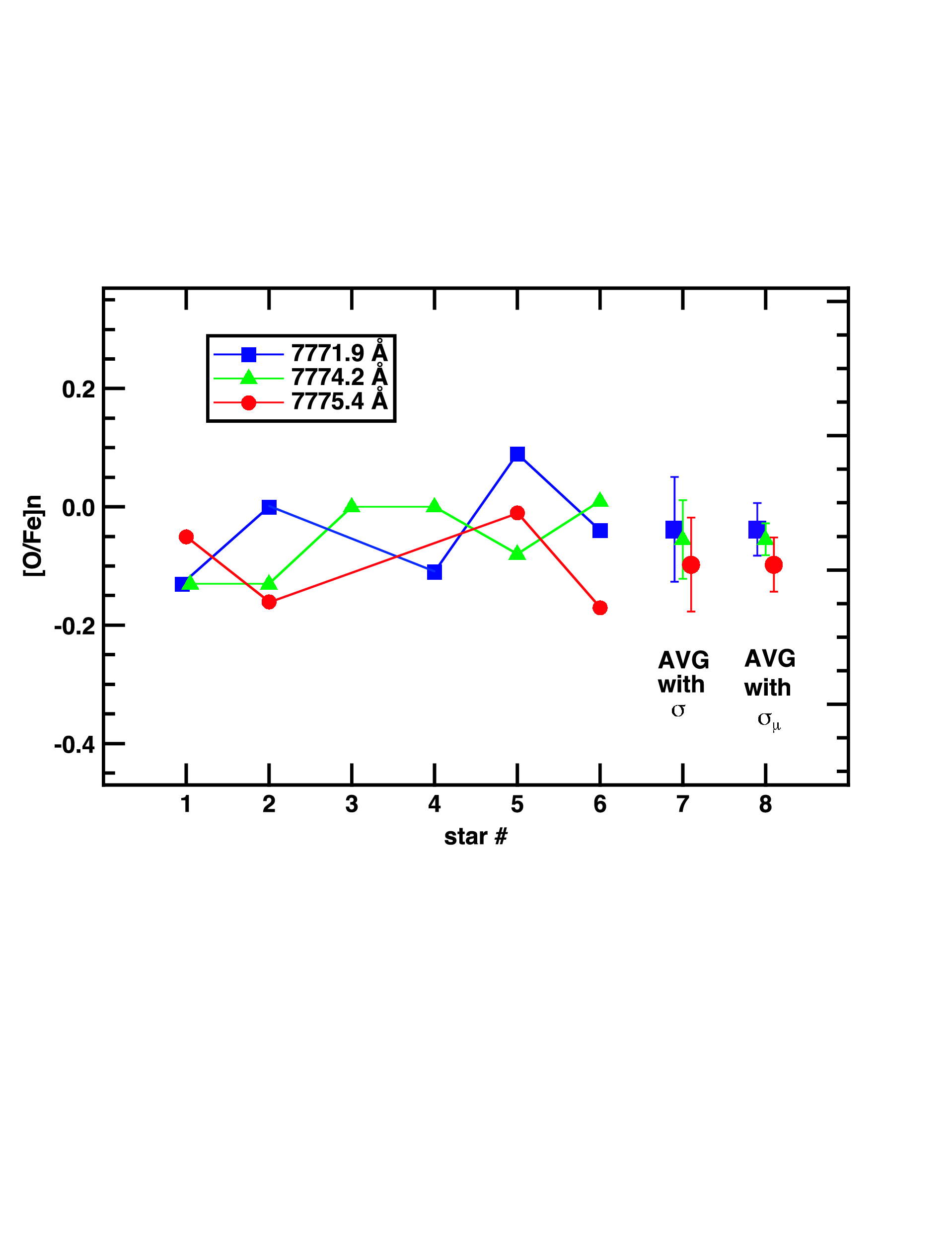}
\caption{Oxygen relative to Fe corrected for nLTE effects for each of the
three individual lines in the O I triplet.  The ``star \#'' corresponds
to increasing MJP numbers as in Table 5.  The filled squares (blue) are from
the 7771.9 line, filled triangles (green) are from the 7774.2 line, and the
filled circles (red) are from 7775.4.  The standard deviation for each line,
$\sigma$, is shown (at \#7) and the standard deviation of the mean,
$\sigma$$_{\mu}$, is also shown (at \#8).  There are no systematic
line-to-line effects and no one star deviates from the others.}
\end{figure}

\begin{figure}
\plotone{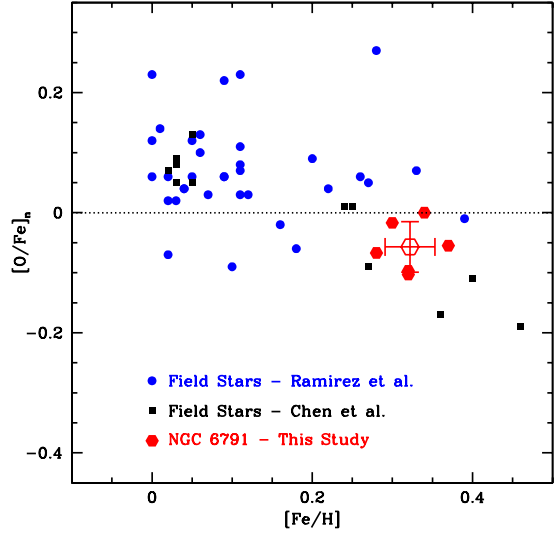}
\caption{Our results for [O/Fe] as corrected for nLTE effects are the (red)
hexagons and are compared with those from Ramirez et al.~(2013) and Chen
et al.~(2003).  The stars selected from Ramirez et al.~are shown as filled
(blue) circles and are metal-rich ([Fe/H] $>$0.00) and old (age $>$7.7 Gyr).
The old (8--10 Gyr) metal-rich ([Fe/H] $>$ 0.01) field stars from Chen et
al.~(2003) are filled squares (black).  The turn-off stars in NGC 6791 have
values of [O/Fe]$_{\rm n}$ that are lower than solar [O/Fe]$_{\rm n}$.  The
trend of decreasing [O/Fe] with increasing [Fe/H] can be seen.}
\end{figure}

\begin{figure}
\plotone{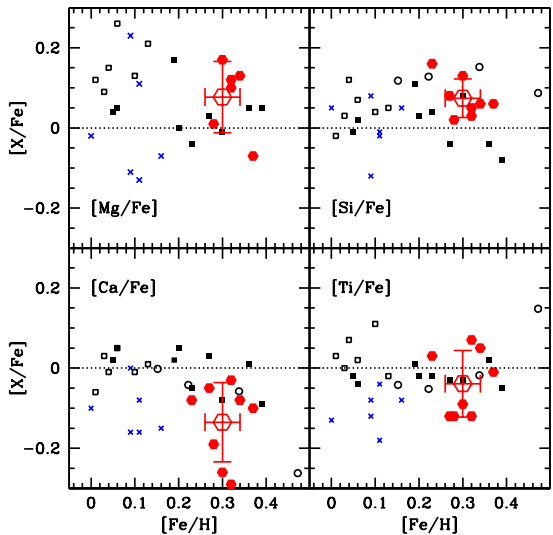}
\caption{Abundances of the four alpha-elements relative to Fe shown as
filled (red) hexagons.  The cluster mean with error bars are shown as open
hexagons.  The comparison field stars are all old ($>$8 Gyr) and metal rich.
The filled squares are from Chen et al.~(2003), the open squares from
Edvardsson et al.~(1993), the open circles from Feltzing \& Gonzalez (2001),
the crosses (blue) are from Reddy et al.~(2006) as selected by age and [Fe/H]
from Ramirez et al.~(2012).}
\end{figure}

\begin{figure}
\plotone{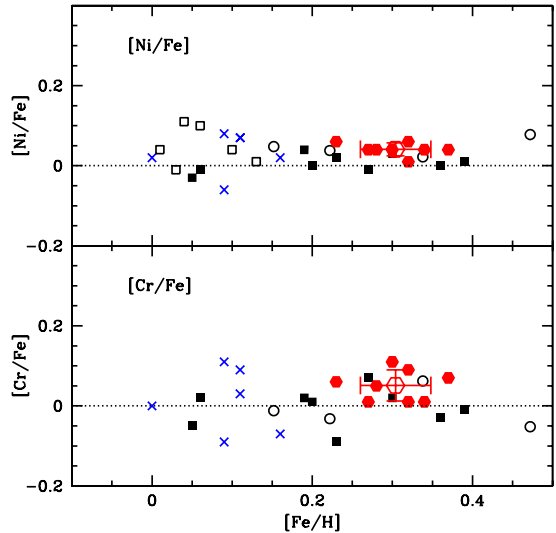}
\caption{Abundances of the Fe-peak elements relative to Fe shown as
filled (red) hexagons.  The cluster mean with error bars are shown as open
hexagons.  The comparison samples are the same as in Figure 6.}
\end{figure}

\begin{figure}
\plotone{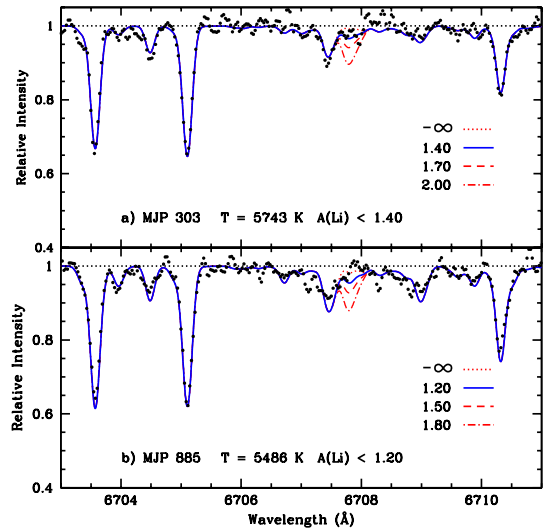}
\caption{Synthesis in the Li region in two of our stars.  The black dots are
the observations.  The solid (blue) line shows the fit to the spectrum and our
estimate of the upper limit on A(Li).  MJP 885 appears to have no Li at all,
but due to the noise in the spectrum, we take A(Li) $<$ 1.2. }
\end{figure}

\begin{figure}
\plotone{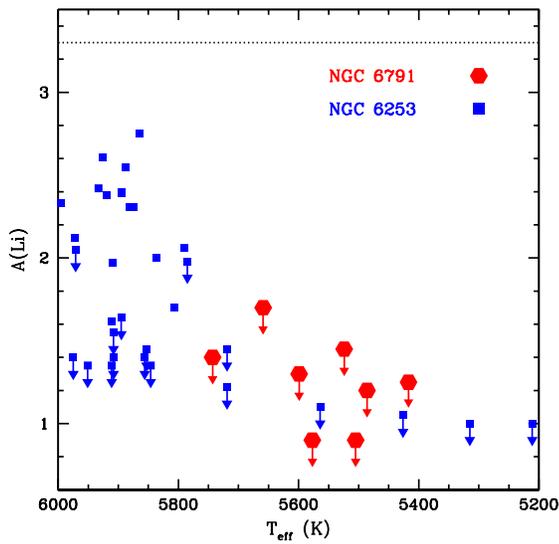}
\caption{Lithium upper limit abundances in NGC 6791 turn-off stars compared
with Li in turn-off stars in the metal-rich cluster NGC 6253 (Cummings et
al.~2012).  The filled hexagons (red) show the upper limit values for the
turn-off stars in NGC 6791.  The filled squares (blue) are the results for NGC
6253 from Cummings et al.~(2012) The downward arrows indicate upper limits in
both clusters.  The horizontal dotted line corresponds to the Li abundance in
meteorites.}
\end{figure}

\begin{figure}
\plotone{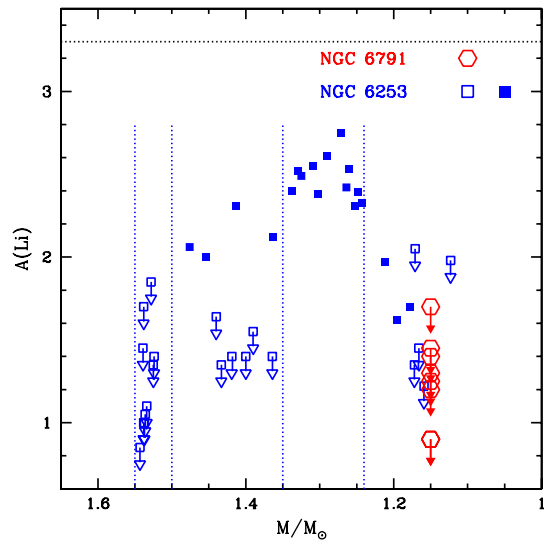}
\caption{Li abundances and upper limits for the two clusters, NGC 6791 and NGC
6253 as a function of stellar mass.  The upper limits for NGC 6791 are shown
as open hexagons with downward arrows; those for NGC 6253 are open squares
with downward arrows.  The Li abundances for NGC 6253 are filled squares.  The
vertical dotted lines separate the particular mass regions and their
evolutionary states as described in the text.  The horizontal dotted line
corresponds to the Li abundance in meteorites.}
\end{figure}


\singlespace
\begin{center}
\begin{deluxetable}{rrcclcc} 
\tablewidth{0pc}
\tablecolumns{7} 
\tablecaption{Keck HIRES Observations} 
\tablehead{ 
\colhead{MJP} &  \colhead{SBG} &  \colhead{V} & \colhead{B-V} &   
\colhead{Night} & \colhead{Exp.(min)} & \colhead {S/N} 
}
\startdata 
 303 & 15592 & 17.426 & 0.843 & 2008 Jul 22  & 6 x 30 & 40 \\
 885 & 14416 & 17.437 & 0.934 & 2009 Aug 04  & 5 x 30 & 40 \\
1328 & 13334 & 17.382 & 0.927 & 2008 Jul 22  & 2 x 45 \\
     &       &        &       & 2008 Aug 14  & 4 x 30 & 37 \\
1346 & 13352 & 17.508 & 0.872 & 2009 Jul 17  & 3 x 30 \\
     &       &        &       & 2009 Aug 04  & 3 x 30 & 30 \\
1783 & 12181 & 17.408 & 0.901 & 2009 Aug 04  & 5 x 30 & 42 \\
2279 & 11220 & 17.366 & 0.920 & 2008 Jul 22  & 3 x 30, 2 x 45 & 37 \\
4112 &  7649 & 17.418 & 0.893 & 2000 May 28  & 2 x 60 \\
     &       &        &       & 2000 May 29  & 2 x 60 & 38 \\
5061 &  5744 & 17.366 & 0.960 & 1999 Jun 07  & 2 x 40 \\
     &	     &	      &       & 1999 Jun 08  & 2 x 50 & 40 \\
5597 &  4591 & 17.384 & 0.829 & 2009 Jul 17  & 6 x 30 & 35 \\
6930 &  2130 & 17.365 & 0.921 & 2009 Jul 17  & 5 x 30 & 37 \\
\enddata

\end{deluxetable}
\end{center}

\clearpage

\singlespace
\begin{center}
\begin{deluxetable}{rrrcccccccc} 
\tablewidth{0pc}
\tablenum{2}
\tablecolumns{11} 
\tablecaption{Colors, Temperatures, Gravities, Microturbulent Velocities} 
\tablehead{ 
\colhead{MJP} &  \colhead{SBG} & \colhead{Bro} & \colhead{$B$}
& \colhead{$V$}
 & \colhead{$\delta_{redd}$} & \colhead{$B-V$} & \colhead{$(B-V)_{0}$} &
\colhead{$T_{\rm eff}$} & \colhead{$log$ g} & \colhead{$\xi$} 
}
\startdata 
303  & 15592 & 28222  & 18.269 & 17.426 &\nodata & 0.843 & 0.688 & 5743 & 4.12   & 1.52 \\
885  & 14416 & 26107  & 18.320 & 17.398 & +0.012   & 0.934 & 0.779 & 5486 & 4.03 & 1.45 \\
1328 & 13334 & 24514  & 18.292 & 17.369 & +0.004   & 0.927 & 0.772 & 5505 & 4.01 & 1.49 \\
1346 & 13352 & 24536  & 18.294 & 17.442 & +0.020   & 0.872 & 0.717 & 5659 & 4.12 & 1.47 \\
1783 & 12181 & 23028  & 18.307 & 17.406 & +0.000   & 0.901 & 0.746 & 5577 & 4.05 & 1.50 \\
2279 & 11220 & 21859  & 18.276 & 17.358 & +0.002   & 0.920 & 0.765 & 5524 & 4.01 & 1.50 \\
4112 & 7649  & 17730  & 18.336 & 17.438 & $-$0.005 & 0.893 & 0.738 & 5599 & 4.06 & 1.50 \\
5061 & 5744  & 15480  & 18.350 & 17.385 & $-$0.005 & 0.960 & 0.805 & 5417 & 3.97 & 1.48 \\
5597 & 4591  & 14096  & 18.297 & 17.449 & $-$0.019 & 0.829 & 0.674 & 5784 & 4.12 & 1.55 \\
6930 & 2130  &\nodata & 18.286 & 17.365 &\nodata & 0.921 & 0.766 & 5522 & 4.01   & 1.47 \\
\enddata
\end{deluxetable}
\end{center}

\clearpage

\tightenlines

\singlespace
\begin{center}
\begin{deluxetable}{lccccc} 
\tablewidth{0pc}
\tablenum{3}
\tablecolumns{6} 
\tablecaption{Spectral Lines Used for Fe Abundances and Equivalent Widths
Measured in
MJP 303 and MJP 2279}  
\tablehead{ 
\colhead{Ion} & \colhead{$\lambda$ (\AA) } & \colhead{Ex.~Pot.~(eV)} &
\colhead{$\log{gf}$} & \colhead{W(m\AA) 303} & \colhead{W(m\AA) 2279}
} 
\startdata 
{\ion{Fe}{1}} & 5560.21 & 4.43 & $-$1.190 & 80.5 & 67.2 \\
&   5775.08 &     4.22 &   $-$1.300 & 81.6 & 80.7 \\
&   6027.05 &	  4.08 &   $-$1.150 & 83.3 & \nodata   \\
&   6082.72 &     2.22 &   $-$3.533 & 61.8 & 70.4   \\
&   6127.90 &	  4.14 &   $-$1.399 & 65.8 & 72.2   \\
&   6151.62 &	  2.18 &   $-$3.299 & 70.9 & 78.0   \\
&   6157.73 &	  4.07 &   $-$1.270 & 85.5 & \nodata   \\
&   6165.36 &     4.14 &   $-$1.470 & 64.1 & 70.6   \\ 
&   6180.20 &     2.73 &   $-$2.622 & 83.5 & 83.5   \\
&   6187.99 &     3.94 &   $-$1.570 & 73.4 & 71.0   \\  
&   6380.74 &     4.19 &   $-$1.388 & 72.6 & 85.4   \\
&   6469.19 &     4.83 &   $-$0.620 & 84.6 & 83.5   \\
&   6475.63 &     2.56 &   $-$2.970 & 80.1 & 84.4   \\
&   6481.87 &	  2.28 &   $-$2.980 & 92.9 & 90.5   \\
&   6496.47 &     4.79 &   $-$0.650 & 82.8 & 86.5   \\
&   6498.94 &	  0.96 &   $-$4.690 & 75.3 & 81.1   \\
&   6581.22 &	  1.48 &   $-$4.790 & 48.7 & 55.2   \\
&   6597.56 &     4.79 &   $-$0.920 & 56.8 & 65.9   \\
&   6608.04 &	  2.28 &   $-$4.020 & 38.0 & 39.5   \\
&   6609.11 &     2.56 &   $-$2.677 & 93.9 & 98.0   \\
&   6627.56 &	  4.55 &   $-$1.610 & 46.2 & 57.0   \\
&   6703.57 &	  2.76 &   $-$3.130 & 62.3 & 61.7   \\
&   6710.32 &     1.48 &   $-$4.900 & 38.3 & 37.8   \\
&   6725.36 &     4.10 &   $-$2.300 & 38.0 & 39.8   \\
&   6726.67 &	  4.61 &   $-$1.160 & 68.0 & 73.0   \\
&   6733.15 &     4.64 &   $-$1.520 & 47.3 & 45.8   \\
&   6739.52 &     1.56 &   $-$4.979 & 26.8 & 40.6   \\
&   6752.71 &     4.64 &   $-$1.200 & 59.2 & 69.5   \\
&   7000.62 &     4.14 &   $-$2.240 & 44.9 & 31.1   \\
&   7068.42 &     4.06 &   $-$1.430 & 90.8 & \nodata   \\
&   7107.47 &	  4.19 &   $-$2.070 & 47.4 & 50.8   \\
&   7127.57 &	  4.99 &   $-$1.250 & 42.8 & 57.0   \\
&   7132.98 &	  4.07 &   $-$1.770 & 58.8 & 82.7   \\
&   7142.52 &	  4.95 &   $-$1.090 & 50.6 & 60.9   \\
&   7155.63 &	  5.01 &   $-$1.090 & 65.7 & 68.9   \\
&   7219.68 &     4.07 &   $-$1.690 & \nodata & 77.3 \\
&   7401.69 &     4.19 &   $-$1.690 & 61.8 & 72.0   \\
&   7568.91 &     4.28 &   $-$0.990 & 94.1 & 95.9   \\
&   7710.37 &     4.22 &   $-$1.220 & 97.1 & 92.5   \\   
&   7723.21 &     2.28 &   $-$3.590 & 71.5 & 85.1   \\ 
&   7745.52 &	  5.08 &   $-$1.180 & 34.4 & 52.2   \\
&   7746.60 &	  5.06 &   $-$1.290 & 38.5 & 40.3   \\
&   7855.40 &     5.06 &   $-$1.200 & 48.1 & 47.8   \\
&   7879.78 &	  5.03 &   $-$1.650 & 21.4 & 19.5   \\
&   7912.87 &     0.86 &   $-$4.900 & \nodata & 97.2 \\
\hline
{\ion{Fe}{2}} & 5534.847 &  3.25  &  $-$2.911 & 68.7 & 72.0  \\
&   6084.10 &     3.20 &   $-$3.860  & 31.2 & 41.8   \\
&   6149.25 &     3.89 &   $-$2.724  & 55.3 & 54.1   \\
&   6247.56 &	  3.89 &   $-$2.329  & \nodata & 65.6   \\   
&   6369.46 &     2.89 &   $-$4.160  & 38.8 & 34.7   \\
&   6456.39 &	  3.90 &   $-$2.075  & 90.1 & \nodata   \\   
&   6516.08 &	  2.89 &   $-$3.380  & 70.7 & 65.5   \\   
&   7224.46 &	  3.89 &   $-$3.243  & 41.1 & \nodata   \\   
&   7711.73 &	  3.90 &   $-$2.450  & 75.0 & \nodata   \\
\enddata

\end{deluxetable} 
\end{center}


\clearpage


\singlespace
\begin{center}
\begin{deluxetable}{lccccccc}
\tablenum{4}
\tablewidth{0pc}
\tablecolumns{8} 
\tablecaption{[Fe/H] Abundances from Fe I and Fe II}
\tablehead{
\colhead{Star}  & \colhead{[FeI/H]} &  \colhead{$\sigma$} & \colhead{num.} & 
\colhead{[FeII/H]} &  \colhead{$\sigma$} & \colhead{num.} & 
\colhead{mean [Fe/H]} 
}
\startdata 
303  & +0.33  & $\pm$0.12  & 42 & +0.27 & $\pm$0.12 & 8 & +0.32 \\
885  & +0.32  & $\pm$0.12  & 37 & +0.30 & $\pm$0.12 & 7 & +0.32 \\
1328 & +0.33  & $\pm$0.12  & 28 & +0.41 & $\pm$0.12 & 6 & +0.34 \\
1346 & +0.37  & $\pm$0.12  & 39 & +0.34 & $\pm$0.12 & 5 & +0.37 \\
1783 & +0.30  & $\pm$0.14  & 40 & +0.30 & $\pm$0.12 & 5 & +0.30 \\
2279 & +0.28  & $\pm$0.14  & 42 & +0.26 & $\pm$0.15 & 6 & +0.28 \\
4112 & +0.27  & $\pm$0.12  & 26 & +0.28 & $\pm$0.06 & 5 & +0.27 \\
5061 & +0.23  & $\pm$0.09  & 26 & +0.24 & $\pm$0.08 & 6 & +0.23 \\
mean & +0.30  & $\pm$0.04 ($\pm$0.02)  & \nodata & +0.30  & $\pm$0.05 ($\pm$0.02) & \nodata & +0.30 $\pm$0.02\\
\hline
5597 & $-$0.05 & $\pm$0.11  & 46 & \nodata & \nodata & \nodata & \nodata \\   
6930 & $-$0.33 & $\pm$0.11  & 41 & \nodata & \nodata & \nodata & \nodata \\   
\enddata 

\end{deluxetable} 
\end{center}


\clearpage


\singlespace
\begin{center}
\begin{deluxetable}{lccccccc}
\tablenum{5}
\tablewidth{0pc}
\tablecolumns{8}
\tablecaption{Oxygen Abundances}
\tablehead{
\colhead{Star}  
& \colhead{$\lambda$(\AA)} 
& \colhead{EQW} & log N(O) & \colhead{corr.}
& \colhead{log N(O)n}  
& \colhead{[O/H]n}
   & \colhead{[O/Fe]n}   
}
\startdata
303    & 7771.94 & 89.8 & 9.02 & $-$0.18 & 8.84 & 0.19 & $-$0.13 \\
+0.32  & 7774.17 & 77.5 & 8.99 & $-$0.15 & 8.84 & 0.19 & $-$0.13 \\
       & 7775.39 & 67.4 & 9.05 & $-$0.13 & 8.92 & 0.27 & $-$0.05 \\
mean   &         &      &      &         &  &    &  $-$0.10 $\pm$0.03\\ 
\hline
885    & 7771.94 & 79.8 & 9.12 & $-$0.15 & 8.97 & 0.32 & \phn0.00 \\   
+0.32  & 7774.17 & 60.7 & 8.96 & $-$0.12 & 8.84 & 0.19 & $-$0.13 \\
       & 7775.39 & 45.6 & 8.90 & $-$0.09 & 8.81 & 0.16 & $-$0.16 \\
mean   &         &      &       &      &   &     & $-$0.10 $\pm$0.06\\
\hline
1328   & 7771.94 &\nodata&\nodata&\nodata  &\nodata &\nodata&\nodata \\ 
+0.34  & 7774.17 & 73.9 & 9.14 & $-$0.14 & 9.00 & 0.34 & \phn0.00 \\
       & 7775.39 &\nodata&\nodata&\nodata  &\nodata&\nodata &\nodata \\
mean   &         &        &      &        &        &       & \phn0.00 \\
\hline
1346   & 7771.94 & 86.9 & 9.08 & $-$0.17  & 8.91 & 0.26 & $-$0.11 \\
+0.37  & 7774.17 & 84.2 & 9.19 & $-$0.17  & 9.02 & 0.37  & \phn0.00 \\
       & 7775.39 &\nodata&\nodata&\nodata &\nodata  &\nodata \\
mean &       &         &      &       &         &         & $-$0.06 $\pm$0.08\\
\hline
1783   & 7771.94 & 89.4 & 9.17 & $-$0.18 & 8.99 & 0.34 & \phn0.04 \\
+0.30  & 7774.17 & 68.6 & 9.01 & $-$0.14 & 8.87 & 0.22 & $-$0.08 \\
       & 7775.39 & 57.8 & 9.05 & $-$0.11 & 8.94 & 0.29 & $-$0.01 \\
mean   &         &      &      &         &      &      & $-$0.02 $\pm$0.04\\ 
\hline
2279   & 7771.94 & 78.0 & 9.04 & $-$0.15 & 8.89 & 0.24 & $-$0.04  \\
+0.28  & 7774.17 & 71.2 & 9.08 & $-$0.14 & 8.94 & 0.29 & \phn0.01 \\
       & 7775.39 & 47.0 & 8.87 & $-$0.10 & 8.76 & 0.11 & $-$0.17 \\
mean   &         &      &      &         &      &      & $-$0.07 $\pm$0.07\\  
\hline
Sun    & 7771.94 & 62.8 & 8.68 & $-$0.11 & 8.57	&\nodata & \nodata\\
0.00   & 7774.17 & 59.8 & 8.78 & $-$0.10 & 8.68	&\nodata & \nodata\\
       & 7775.39 & 46.6 & 8.77 & $-$0.07 & 8.70	&\nodata & \nodata\\
mean   &         &      & 8.74 &         & 8.65 &	 & \\
\enddata 

\end{deluxetable} 
\end{center}


\clearpage


\singlespace
\begin{center}
\begin{deluxetable}{rrcrcc}
\tablenum{6}
\tablewidth{0pc}
\tablecolumns{6}
\tablecaption{Comparison Field Stars for Oxygen}
\tablehead{
\colhead{HIP} & \colhead{HD}  & \colhead{[Fe/H]}  & \colhead{Gyr} 
  & \colhead{[O/H]n}    & \colhead{[O/Fe]n}   
}
\startdata

10303  & 13612    &  0.10 &  8.24  &  +0.01    &  $-$0.09  \\ 
11505  & 15069    &  0.02 &  9.01  &  +0.08    &  +0.06    \\ 
16467  & 21727    &  0.02 &  9.01  &  $-$0.09  &  $-$0.07  \\ 
21703  & 29528    &  0.16 &  8.67  &  +0.14    &  $-$0.02  \\ 
25052  & 34634    &  0.09 &  8.28  &  +0.15    &  +0.06    \\ 
31419  & 47051    &  0.06 &  9.39  &  +0.19    &  +0.13    \\ 
33382  & 51219    &  0.00 &  9.21  &  +0.06    &  +0.06    \\ 
36210  & 54468    &  0.03 &  8.28  &  +0.05    &  +0.02    \\ 
43587  & 75732    &  0.39 &  8.67  &  +0.38    &  $-$0.01  \\ 
55288B & \nodata  &  0.05 &  8.91  &  +0.17    &  +0.12    \\ 
57532  & 102326   &  0.22 &  8.67  &  +0.26    &  +0.04    \\ 
58576  & 104304   &  0.26 &  7.93  &  +0.32    &  +0.06    \\ 
62039  & 110537   &  0.11 &  8.45  &  +0.22    &  +0.11    \\ 
62198  & 110885   &  0.07 &  8.60  &  +0.10    &  +0.03    \\ 
63881  & 113712   &  0.12 &  8.66  &  +0.15    &  +0.03    \\ 
65049  & 115968   &  0.20 &  10.62 &  +0.29    &  +0.09    \\ 
66514  & 118742   &  0.00 &  11.45 &  +0.12    &  +0.12    \\ 
68184  & 122064   &  0.11 &  13.89 &  +0.34    &  +0.23    \\ 
69972  & 125072   &  0.28 &  8.76  &  +0.55    &  +0.27    \\ 
72848  & 131511   &  0.04 &  10.19 &  +0.08    &  +0.04    \\ 
73078  & 132130   &  0.05 &  9.69  &  +0.11    &  +0.06    \\ 
73815  & 133600   &  0.02 &  8.15  &  +0.04    &  +0.02    \\ 
81748  & 150633   &  0.11 &  9.79  &  +0.19    &  +0.08    \\ 
82265  & 151504   &  0.11 &  10.16 &  +0.18    &  +0.07    \\ 
82750  & 153344   &  0.18 &  9.20  &  +0.12    &  $-$0.06  \\ 
86796  & 160691   &  0.27 &  8.18  &  +0.32    &  +0.05    \\ 
94615  & 230999   &  0.09 &  11.65 &  +0.15    &  +0.06    \\ 
96901  & 186427   &  0.04 &  8.91  &  +0.08    &  +0.04    \\ 
99240  & 190248   &  0.33 &  10.65 &  +0.40    &  +0.07    \\ 
99825  & 192310   &  0.06 &  9.41  &  +0.16    &  +0.10    \\ 
102081 & 197207   &  0.09 &  10.61 &  +0.31    &  +0.22    \\ 
106678 & 205656   &  0.01 &  10.30 &  +0.15    &  +0.14    \\ 
114622 & 219134   &  0.00 &  10.93 &  +0.23    &  +0.23    \\ 
115066 & 219781   &  0.11 &  8.29  &  +0.14    &  +0.03    \\ 
\enddata 

\end{deluxetable} 
\end{center}


\clearpage


\singlespace
\begin{center}
\begin{deluxetable}{rcccccccc}
\tablenum{7}
\tablewidth{0pc}
\tablecolumns{9}
\tablecaption{Comparison Field Stars for Mg, Si, Ca, Ti, Cr, Ni}
\tablehead{
\colhead{HD}  & \colhead{[Fe/H]}  & \colhead{[Mg/Fe]} & \colhead{[Si/Fe]}
& \colhead{[Ca/Fe]} & \colhead{[Ti/Fe]} & \colhead{[Cr/Fe]} &
\colhead{[Ni/Fe]} & \colhead{ref}\tablenotemark{1}
}
\startdata
1461    &  +0.39 & +0.05  &  $-$0.08 & $-$0.09 & $-$0.05 & $-$0.01 & +0.01   & C \\
73393   &  +0.05 & +0.04  &  $-$0.01 & +0.02   & $-$0.02 & $-$0.05 & $-$0.03 & C \\
86728   &  +0.20 & +0.00  &  +0.03   & +0.05   & $-$0.02 & +0.01   & \phn0.00    & C \\
104304  &  +0.30 & $-$0.01 &  +0.08  & $-$0.08 & $-$0.03 & +0.02   & +0.03   & C \\
127334  &  +0.27 & +0.03  &  $-$0.04 & +0.03   & $-$0.03 & +0.07   & $-$0.01 & C \\
175518  &  +0.36 & +0.05  &  $-$0.04 & +0.01   & +0.02	 & $-$0.03 & \phn0.00    & C \\
182488  &  +0.23 & $-$0.04 &  +0.04  & $-$0.05 & $-$0.02 & $-$0.09 & +0.02   & C \\
186408  &  +0.06 & +0.05  &  +0.02   & +0.05   & $-$0.04 & +0.02   & $-$0.01 & C \\
190360  &  +0.19 & +0.17  &  +0.11   & +0.02   & +0.01	 & +0.02   & +0.04   & C \\
\hline
45701  &  +0.13 & +0.21 & +0.05   & +0.01   & $-$0.02 & \nodata& +0.01	& E \\
67228  &  +0.04 & +0.15 & +0.12   & $-$0.01 & +0.07    & \nodata& +0.11	& E \\
76151  &  +0.01 & +0.12 & $-$0.02 & $-$0.06 & +0.03    & \nodata& +0.04	& E \\
108309 &  +0.10 & +0.13 & +0.04   & $-$0.01 & +0.11    & \nodata& +0.04	& E \\
177565 &  +0.03 & +0.09 & +0.03   & +0.03   & \phn0.00 & \nodata& $-$0.01& E \\
217014 &  +0.06 & +0.26 & +0.07   & +0.05   & +0.02    & \nodata& +0.10	& E \\
\hline	
99491   & +0.22 &\nodata& +0.13 & $-$0.04 & $-$0.05 & $-$0.05 & +0.04 & FG \\
104304  & +0.15 &\nodata& +0.12 & \phn0.00 & $-$0.04 & $-$0.04 & +0.05 & FG \\
145675  & +0.47 &\nodata& +0.09 & $-$0.26 & +0.15   & +0.15   & +0.08 & FG \\
182572  & +0.34 &\nodata& +0.15 & $-$0.06 & $-$0.02 & $-$0.02 & +0.02 & FG \\
\hline
21703   & +0.16 & +0.16   & +0.05   & $-$0.15 & $-$0.08  & $-$0.07  & +0.02  & R \\
33382   & +0.00 & $-$0.06 & +0.05   & $-$0.10 & $-$0.13  & \phn0.00 & +0.02  & R \\
81748   & +0.11 & +0.13   & $-$0.02 & $-$0.16 & $-$0.18  & +0.09    & +0.07  & R \\
82265   & +0.11 & +0.02   & $-$0.01 & $-$0.08 & $-$0.04  & +0.03    & +0.07  & R \\
94615   & +0.09 & $-$0.17 & $-$0.12 & $-$0.16 & $-$0.12  & +0.11    & +0.08  & R \\
102081  & +0.09 & +0.23   & \phn0.08 &\phn0.00 & $-$0.08 & $-$0.09 & $-$0.06 & R \\ 
\enddata 
\tablenotetext{1}{References: C = Chen et al.~(2003), 
E = Edvardsson et al.~(1993),
FG= Feltzing \& Gonzalez (2001), R = Reddy et al.~(2006)}
\end{deluxetable} 
\end{center}


\clearpage


\singlespace
\begin{center}
\begin{deluxetable}{lccccccccc}
\tablenum{8}
\tablewidth{0pc}
\tablecolumns{10} 
\tablecaption{Elemental Abundances}
\tablehead{
\colhead{Star}  & \colhead{[Fe/H]}   & \colhead{A(Li)}    
& \colhead{[O/Fe]$_n$}
   & \colhead{[Mg/Fe]} & \colhead{[Si/Fe]} & \colhead{[Ca/Fe]} &
  \colhead{[Ti/Fe]} & \colhead{[Cr/Fe]} & \colhead{[Ni/Fe]} }
\startdata 
303  & +0.32 & $<$1.40 & $-$0.10 & +0.10   & +0.03 & $-$0.03 & +0.07    & 0.09 & 0.06   \\
885  & +0.32 & $<$1.20 & $-$0.10 & +0.12   & +0.05 & $-$0.29 & $-$0.12  & 0.01 & 0.01   \\
1328 & +0.34 & $<$1.40 & \phn0.00 & +0.13   & +0.06 & $-$0.08 & $-$0.07  & 0.01  &0.04	\\
1346 & +0.37 & $<$1.70 & $-$0.06 & $-$0.07 & +0.06 & $-$0.10 & $-$0.01  & 0.07  &0.04	\\
1783 & +0.30 & $<$0.90 & $-$0.02 & +0.17   & +0.13 & $-$0.26 & $-$0.09  & 0.11  &0.04	\\
2279 & +0.28 & $<$1.45 & $-$0.07 & +0.01   & +0.02 & $-$0.19 & $-$0.09  & 0.05  &0.04	\\
4112 & +0.27 & $<$1.30 &\nodata  &\nodata  & +0.08 & $-$0.05 & $-$0.12  & 0.01  &0.04	\\
5061 & +0.23 & $<$1.25 &\nodata  &\nodata  & +0.16 & $-$0.08 & +0.03    & 0.06  &0.06	\\
\hline
mean & +0.304 & \nodata & $-$0.057 & +0.077   & +0.074 & $-$0.129 & $-$0.050    & 0.045
  & 0.041  \\
$\sigma$&$\pm$0.044& \nodata &$\pm$0.042 & $\pm$0.089 & $\pm$0.048 & $\pm$0.106 &
$\pm$0.071 & $\pm$0.043  & $\pm$0.015 \\ 
$\sigma$$_{\mu}$&$\pm$0.016& \nodata &$\pm$0.019 & $\pm$0.040 & $\pm$0.018 & $\pm$0.040 &
$\pm$0.027 & $\pm$0.016  & $\pm$0.006 \\ 

\enddata

\end{deluxetable} 
\end{center}


\clearpage


\singlespace
\begin{center}
\begin{deluxetable}{lcccccc}
\tablenum{9}
\tablewidth{0pc}
\tablecolumns{7} 
\tablecaption{Comparison of Abundances for Turn-off Stars and Giants in NGC
6791} 
\tablehead{
\colhead{Element} & \colhead{Turn-off Stars} & \colhead{$\sigma$$_{\mu}$} &
\colhead{Carretta Giants}
& \colhead{$\sigma$} & \colhead{Carraro Giants}
& \colhead{$\sigma$}
}
\startdata
$[$Fe/H$]$  & +0.30   & 0.02 & +0.47   & 0.07 &	+0.39   & 0.02 \\
$[$O/Fe$]$  & $-$0.06 & 0.02 & $-$0.31 & 0.08 &	\nodata & \nodata \\ 
$[$Mg/Fe$]$ & +0.08   & 0.04 & +0.20   & 0.05 &	\nodata & \nodata \\
$[$Si/Fe$]$ & +0.07   & 0.02 & $-$0.01 & 0.10 &	+0.02   & 0.03 \\
$[$Ca/Fe$]$ &$-$0.13  & 0.04 & $-$0.15 & 0.08 &	$-$0.03 & 0.02 \\
$[$Ti/Fe$]$ &$-$0.05  & 0.03 & $-$0.03 & 0.09 &	$-$0.02 & 0.03 \\
$[$Ni/Fe$]$ &+0.04    & 0.01 & $-$0.07 & 0.07 &	$-$0.01 & 0.04 \\
\enddata 

\end{deluxetable} 
\end{center}


\clearpage


\singlespace
\begin{center}
\begin{deluxetable}{llcrrrc}
\tablewidth{0pc}
\tablecolumns{7} 
\tablenum{10}
\tablecaption{Errors due to Uncertainties in the Stellar Parameters}
\tablehead{
\multicolumn{1}{c}{Star}  & 
\multicolumn{1}{c}{Element}  & 
\multicolumn{1}{c}{$\Delta$T$_{\rm eff}$} &
\multicolumn{1}{c}{$\Delta$log g} &
\multicolumn{1}{c}{$\Delta$[Fe/H]} & 
\multicolumn{1}{c}{$\Delta\xi$} & 
\multicolumn{1}{c}{Total} \\
\multicolumn{1}{c}{} & 
\multicolumn{1}{c}{} & 
\multicolumn{1}{c}{$\pm$75 K} & 
\multicolumn{1}{c}{$\pm$0.20} & 
\multicolumn{1}{c}{$\pm$0.10} & 
\multicolumn{1}{c}{$\pm$0.20} & 
\multicolumn{1}{c}{}
}
\startdata 
MJP 303&Fe I &	 $\mp$0.05 & $\pm$0.01 &  0.00	    & $\pm$0.04 &  0.06 \\
&	Fe II&	 $\pm$0.04 & $\mp$0.07 &  $\mp$0.02 & $\pm$0.06 &  0.09 \\
&	O I  &	 $\pm$0.07 & $\mp$0.05 &  0.00	    & $\pm$0.03 &  0.09 \\
&	Mg I &	 $\mp$0.03 & $\pm$0.04 &  0.00	    & $\pm$0.02 &  0.05 \\
&	Si I &	 $\mp$0.01 & $\pm$0.02 &  $\mp$0.01 & $\pm$0.03 &  0.04 \\
&	Ca I &	 $\mp$0.06 & $\pm$0.04 &  $\mp$0.01 & $\pm$0.05 &  0.09 \\
&	Ti I &	 $\mp$0.08 & $\pm$0.01 &  0.00	    & $\pm$0.04 &  0.09 \\
&	Cr I &	 $\mp$0.07 & $\pm$0.01 &  0.00	    & $\pm$0.03 &  0.08 \\
&	Ni I &	 $\mp$0.05 & 0.00      &  $\mp$0.01 & $\pm$0.04 &  0.06 \\
\hline							
MJP 885&Fe I &	 $\mp$0.04 & $\pm$0.01 &  0.00	    & $\pm$0.05 & 0.06 \\
&	Fe II&	 $\pm$0.06 & $\mp$0.10 &  $\mp$0.02 & $\pm$0.07 & 0.14 \\
&	O I  &	 $\pm$0.08 & $\mp$0.08 &  0.00	    & $\pm$0.03 & 0.12 \\
&	Mg I &	 $\mp$0.04 & $\pm$0.03 &  $\pm$0.01 & $\pm$0.04 & 0.06 \\
&	Si I &	 $\pm$0.01 & $\pm$0.02 &  $\mp$0.01 & $\pm$0.04 & 0.04 \\
&	Ca I &	 $\mp$0.06 & $\pm$0.04 &  $\mp$0.01 & $\pm$0.06 & 0.09 \\
&	Ti I &	 $\mp$0.08 & $\pm$0.02 &  $\pm$0.01 & $\pm$0.04 & 0.09 \\
&	Cr I &	 $\mp$0.07 & $\pm$0.02 &  0.00	    & $\pm$0.03 & 0.08 \\
&	Ni I &	 $\mp$0.03 & $\mp$0.01 &  $\mp$0.01 & $\pm$0.05 & 0.06 \\
\enddata 

\end{deluxetable} 
\end{center}


\clearpage

\end{document}